\documentclass[flushrt]{aastex}
\usepackage[onecolumn]{emulateapj5}
\received{2000 October 13}
\revised{2001 April 3}
\accepted{2001 April 17}
\journalinfo{Astrophysical Journal, in press.  astro-ph/0105039}
\submitted{Received 2000 October 13; accepted 2001 April 17.}

\usepackage{natbib}
\usepackage{graphicx}
\usepackage{ifthen}
\newcommand{\microns}{${\mu}$m\ }
\newcommand{\micronp}{${\mu}$m}
\newcommand{\degrees}{$^\circ$}

\begin{document}
\bibliographystyle{apj}

\title{Models of the SL9 Impacts II. \\Radiative-hydrodynamic Modeling
of the Plume Splashback}
\shorttitle{SL9 Impact II. Shock Structure and Light Curves}
\author{Drake Deming}
\affil{Planetary Systems Branch, Code 693}
\affil{Goddard Space Flight Center, Greenbelt, Maryland 20771}
\and
\author{Joseph Harrington}
\affil{Department of Astronomy}
\affil{Cornell University, Ithaca, NY 14853}
\shortauthors{Deming and Harrington}

\begin{abstract}

We model the plume `splashback' phase of the Shoemaker-Levy~9 (SL9)
collisions with Jupiter.  We modified the ZEUS-3D hydrodynamic code to
include radiative transport in the gray approximation, and present
validation tests.  After initializing with a model Jovian atmosphere,
we couple mass and momentum fluxes of SL9 plume material, as
calculated by the ballistic Monte-Carlo plume model of Paper I of this
series.  A strong and complex shock structure results.  The shock
temperatures produced by the model agree well with observations, and
the structure and evolution of the modeled shocks account for the
appearance of high excitation molecular line emission after the
peak of the continuum light curve.  The splashback region cools by
radial expansion as well as by radiation.  The morphology of our
synthetic continuum light curves agree with observations over a broad
wavelength range (0.9--12 {\micronp}).  Much of the complex structure
of these light curves is a natural consequence of the temperature
dependence of the Planck function, and the plume velocity
distribution.  A feature of our ballistic plume is a shell of mass at
the highest velocities, which we term the `vanguard'.  Portions of the
vanguard ejected on shallow trajectories produce a lateral shock
front, whose initial expansion accounts for the `third precursors'
seen in the 2-{\microns} light curves of the larger impacts, and for
hot methane emission at early times observed by \citeauthor{dinelli97}
Continued propagation of this lateral shock approximately reproduces
the radii, propagation speed, and centroid positions of the large
rings observed at 3--4 {\microns} by \citeauthor{mcgregor96}
The portion of the vanguard ejected closer to the vertical falls back
with high $z$-component velocities just after maximum light, producing
CO emission and the `flare' seen at 0.9 {\micronp}.  The model also
produces secondary maxima (`bounces') whose amplitudes and periods are
in agreement with observations.

\end{abstract}

\section{Introduction}
\label{secintro}

In this paper we model the collapse of a nominal ejecta plume from the
collision of comet Shoemaker-Levy 9 (SL9) with Jupiter. We couple the
mass and momentum fluxes from the ballistic plume model of
\citet[hereafter Paper I]{paper1} to a radiative hydrodynamic model of
the jovian atmosphere, and we follow the shock formation, evolution,
and cooling in detail. The ballistic plume model was developed in
Paper I based on the power-law velocity distribution of \citet{ZM94},
with free parameters constrained to reproduce the appearance of the
observed plumes at the jovian limb and the disk debris patterns as
observed by \citet{hammel95}. In the present model, only a few additional
parameters can be varied, and we have simply adopted reasonable
values, without attempting to `fine-tune' the model's agreement with
observations.  It is not our intent to `fit' the model to the
observations for specific impacts, but merely to elucidate the physics
of a typical splashback.  Nevertheless, the success of our approach is
exemplified by good agreement between our synthetic light curves and
observations over a broad wavelength range (0.9 -- 12 {\micronp}), and
by the occurrence in the model output of several phenomena which have
previously eluded explanation.

We describe the physical basis of the radiative hydrodynamic model in
Sec.\ref{seccp} and its validation using test problems in
Sec.\ref{seccv}.  In Sec.\ref{secmr}, we apply it to the nominal SL9
plume, and compare the results to observations.   Section
\ref{secsum} summarizes our results.

\section{Code Physics}
\label{seccp}

The radiative hydrodynamic model utilizes the ZEUS-3D hydrocode
\citep{stone92,clarke94}, which we have modified to
include radiative transport in the gray approximation.  The physics
in our modified code is given by the fluid equations:
\newpage
\begin{eqnarray}
\textrm{Continuity:} & \frac{\partial \rho}{\partial t} + \nabla\cdot (\rho\mathbf{v}) = 0\label{eqcont}\\
\textrm{Momentum:} & \frac{\partial\mathbf{P}}{\partial t} + \nabla\cdot
(\mathbf{Pv}) = - \nabla p - \rho \mathbf{g}\label{eqmom}\\
\textrm{Energy:} & \frac{\partial e}{\partial t} + \nabla\cdot ( e \mathbf{v}) =
-p\nabla\cdot\mathbf{v} +{4}\pi\kappa\rho (J-S)\label{eqenergy}\\
\textrm{Ideal Gas Equation of State:} & p=(\gamma-1)e=R\rho T\label{eqidgas},
\end{eqnarray}
with mass density $\rho$, time $t$, velocity flow field
$\mathbf{v}$, momentum vector field $\mathbf{
P}=\rho\mathbf{v}$, pressure $p$, (constant) gravitational
acceleration vector $\mathbf{g}$, internal energy per unit volume
$e$, opacity $\kappa$, mean intensity $J$, source function
$S$, the (constant) ratio of specific heats $\gamma$, gas constant
$R$, and temperature $T$.  We also add the radiative transfer
equations:

\begin{equation}
J(\tau_{0})=\frac{1}{2} \int_0^{\infty} S(\tau)E_1(\tau - \tau_0) \mid
\tau - \tau_0 \mid d\tau\label{eqint}
\end{equation}
\begin{equation}
S=\frac{\sigma T\sp{4}}{\pi}\label{eqsrc}
\end{equation}
\begin{equation}
\tau=\int \kappa \rho\  dz\label{eqodepth}
\end{equation}
with optical depth $\tau$, first exponential
integral function $E_1$ \citep{mihalas78}, Steffan-Boltzmann
constant $\sigma$, and height above the 1-bar level in the jovian
atmosphere $z$.  Note that our implementation of radiative transfer
in ZEUS-3D treats radiation as a {\em transport mechanism}, not as a
{\em fluid}.  We ignore the internal energy and stresses of the
radiation itself, and assume its propagation to be instantaneous.

The model uses quadrature integration to calculate $J$ from Eq.\ \ref{eqint}.
Shocked regions occupying only a few grid zones are adequately
represented in the quadrature, allowing the correct reproduction of
any possible heating of the lower atmosphere by radiation from the
overlying shocks \citep{melosh90}.  Note also that Eq.\ \ref{eqsrc}
represents an assumption of local thermodynamic equilibrium (LTE) for
the radiation field.

The radiative transfer Eqs.\ \ref{eqint}--\ref{eqodepth} require an
additional boundary condition, represented by the effective
temperature of the atmosphere ($T_\mathrm{eff}$) and the consequent
radiative flux which passes continuously through the fluid.  At the
lower boundary of the atmosphere, where $\tau \gg 1$, we utilize the
approximations \citep{mihalas78}:

\begin{equation}
{J-S}  =  \frac{1}{3} \frac{d^2S}{d\tau^2}\label{eqintminsrc}
\end{equation}
\
\begin{equation}
\frac{dS}{d\tau} =  \frac{3 \sigma T^4}{4\pi}\label{eqdsdtau}.
\end{equation}
In order to implement Eqs.\ \ref{eqintminsrc} and \ref{eqdsdtau} in
the code, we add a `virtual' layer below the deepest active layer. We
calculate the optical depth and temperature of the virtual layer in
advance from $T_\mathrm{eff}$, and use them in calculating the
space-centered derivatives of $S$ with respect to $\tau$.  Equation
\ref{eqintminsrc} then gives the value of $J-S$ when the optical depth
exceeds some pre-determined value (typically 10).

ZEUS uses an explicit finite difference method to integrate the fluid
equations, with a Courant condition \citep{courant48} imposed to limit
the size of the time step.  Our addition of radiation to the energy
equation, and the influx of plume mass and momentum to the top layer
of the atmosphere, must also be considered in the context of the ZEUS
finite-difference approximations.  We re-derived the finite difference
approximation to Eq.\ \ref{eqenergy} to maintain correct centering in
space and time, and we found that requiring the change in $e$ to be
less than 10\% from radiation in a single time step gives stable
results.  We similarly require that the plume mass added to the top
atmospheric layer cannot change $\rho$ by more than 100\% in a single
time step.

The mass and momentum fluxes in the ballistic plume are relatively
uniform over a `core' region defined by the opening angle of the
`launch cone' in $r\theta$ coordinates.  ZEUS-3D was therefore
configured to solve Eqs.\ \ref{eqcont}--\ref{eqdsdtau} in $zr\theta$
geometry, where $r$ is radial distance from the impact point and
$\theta$ is the azimuthal angle. Given the approximate uniformity in
$\theta$ for the bulk of the infalling mass, all models presented here
use a single 2-D wedge in $\theta$, which provides for the plume's
vertical and radial velocity components, and permits buoyancy effects
(gravity waves). We neglect Coriolis effects and the curvature of the
jovian atmosphere, which will not significantly affect the shock
structure and light curves.  Figure \ref{figgeom} shows a sketch of
the $zr\theta$ geometry as implemented in ZEUS.
 
\citet[hereafter ZM95]{ZM95} computed a theoretical light curve under
the assumption that the infalling plume energy was instantaneously
balanced by radiative losses.  Since radiative losses have a time
scale which varies as $T^{-3}$, a detailed treatment of radiative
damping is needed.  Radiation in numerous molecular bands and
continuum emission from grains are both potentially important.
However, the detailed wavelength dependence of the radiation is less
important to the hydrodynamics than the bolometric energy loss, so we
use a mean opacity.  Also, since the scale height of the jovian
atmosphere is much less than the radial scale of the plume splashback,
we compute the mean intensity at each point under the assumption that
the atmosphere is homogenous in $r$ and $\theta$.  In other words, the
radiation field does not `see' points at other values of $r$ and
$\theta$, only the variation in the $z$ direction is accounted for
(this is implicit in Eq.\
\ref{eqint}).

A significant portion of the plume mass was likely present in the form
of silicate or other grains
\citep{nicholson95b,griffith97,molina97,friedson98} whose deposition
is believed to account for the post-impact debris patterns on the
jovian disk.  However, \citet{takata97}, \citet{zahnle96}, and Paper I
calculate that the plumes were mostly entrained jovian air.  We assume
that the overall hydrodynamics of the splashback was determined by the
gaseous component, with grains advected by the flow and heated by the
shocks \citep{moses97}.  Nevertheless, we have incorporated an option
in the code for following the motion of tracer grains, by explicit
integration of the grain momentum equations.  Ballistic studies
\citep{jessup00,paper1} have shown conclusively that continued radial
transport after splashback is necessary to explain the disk debris
patterns.  We performed exploratory grain transport calculations,
which indicated good agreement with the observed radial extent of disk
debris, but also convinced us that the transport of grains does not
have a major effect on the phenomena discussed in this paper (e.g.,
light curves).

The original version of ZEUS-3D applies to a single-component fluid,
having a single value for $R$ and $\gamma$.  We adopt $\gamma=1.2$
from the Galileo observations of the G-impact fireball
\citep{carlson97}.  We also apply this value to the jovian
atmosphere, lacking a direct determination for shocked jovian air.
However, we have modified ZEUS to use two values for the gas constant
($R$), consistent with different molecular weights ($\mu$) for the
plume material and for jovian air.  We use $\mu= 2.28$ g\ mole$^{-1}$
for jovian air, which reflects the helium abundance from the Galileo
Probe Helium Interferometer \citep{vonZahn98}.  For the plume, we
calculated $\mu$ assuming that the impacting object was predominately
water ($\mu=18$); the fraction of the plume material which was
impactor was determined in the ballistic plume modeling, and was not a
free parameter in ZEUS-3D.  To explore the `zero-th order' effects of
large versus small impacts, we vary the plume mass by a scale factor,
keeping the same proportions of impactor and jovian air in the plume.

ZEUS-3D uses a source step and a transport step to integrate the fluid
equations \citep{stone92}.  The source step accelerates the fluid
using the pressure gradient and gravity terms on the right of Eq.\
\ref{eqmom} and the radiation term on the right of Eq.\
\ref{eqenergy}.  The transport step integrates the spatial and
temporal gradients to advect the fluid variables across the Eulerian
spatial grid, based on the \citet{vanLeer77} scheme.  Exploratory
calculations indicated that there is little mixing at the
plume/atmosphere interface in the first hours after impact, so we used
the following simplified treatment. We added an additional van Leer
advection calculation to ZEUS-3D, tracking the total overlying column
density of plume material at each $r$, thus determining the location
of the plume/atmosphere boundary.  This permits two values of $R$,
depending on whether the fluid in a given grid zone is plume,
atmosphere, or both.

\section{Code Validation}
\label{seccv}

The ZEUS hydrocode has itself been tested well against a suite of
standard hydrodynamic test problems \citep{stone92}.  Since we have
modified the code, we must establish the validity of our modifications
and also show that we have not inadvertently interfered with the
existing code algorithms.  We have performed numerous tests for these
purposes, including direct checks of some code algorithms against hand
calculations. We also conducted five more-complex tests of the code,
as described below.

\subsection{Test 1: Shock Tubes}  
\label{ssect1st}

The shock tube is a one-dimensional test problem wherein a boundary
separates two isotropic regions of different pressures and densities.
When the boundary is removed, a shock propagates into the lower
pressure region, and the code results can be compared to an analytic
solution \citep{sod78}.  For these tests, we configured ZEUS-3D in a
1-D mode with 1000 zones and with the radiation terms turned off.  We
first repeated the shock tube test described by \citet{stone92}, and
verified that we can reproduce their Fig.\ 11.  However, shocks in the
SL9 problem will have larger compression ratios, e.g., the jump in
pressure across the shock can be as large as $10^{4}$.  Figure
\ref{figtest} shows a successful test for this compression ratio.

\subsection{Test 2: Hydrostatic Equilibrium and Scale Height}
\label{ssect2hesh}

The second term on the right of Eq.\ \ref{eqmom} represents Jupiter's
constant gravitational acceleration. To test the addition of this
term, we started with a 2-D volume with initial constant pressure but
oscillatory temperature and density.  Since constant pressure is not a
solution to (2), the code evolved this atmosphere.  We used the
radiative damping option, with the arbitrary constraint that $J=S$ for
$T=500$ K.  (This is equivalent to placing an optically-thin
atmosphere in a blackbody cavity at this temperature.)  At ${t=8\times
10^5}$ sec, a near-equilibrium state with constant temperature and
gradients in pressure and density prevailed.  We verified that there
were no significant horizontal gradients, and the pressure gradient
term balanced the $\mathbf{g}\rho$ term to within a fractional error
$<{5 \times 10^{-6}}$.  The scale height agreed with the analytic
value to within 0.1\%.

\subsection{Test 3: Radiative Damping Time Scale}
\label{ssect3rdts}

The radiative damping time of a temperature perturbation depends on
its spatial scale as well as its temperature amplitude. Perturbations
of large spatial scale can extend over many optical depths, and these
damp slowly because their optical thickness hinders photon transfer.
\citet{spiegel57} derived an analytic expression for the damping time
of small amplitude temperature perturbations in an isothermal
atmosphere in LTE.  We turned off the gravitational acceleration term
in Eq.\ \ref{eqmom} and configured ZEUS-3D for a 1-D isothermal
atmosphere ($T=5000$ K), with a superposed sinusoidal temperature
perturbation of amplitude 100 K.  These relatively high temperatures
assure that radiative transfer is rapid in comparison to hydrodynamic
effects, which are not included in the analytic damping formula.  We
used 1000 zones having optical thickness of 0.02 per zone.  We ran the
code for times which are small compared to the e-folding time of the
perturbation (so that the fluid was not significantly accelerated),
and we computed the damping rate of the perturbation amplitude,
$\partial(\Delta T) /\partial{t}$, from the model output.  Figure
\ref{figtest} shows this rate compared to the analytic solution
as a function of the wavelength of the perturbation, measured in
optical depths, $\tau_{\lambda}$. The largest difference between the
code and the analytic formula is 6\% at $\tau_{\lambda} = 2$. Since
$S$ represents a small perturbation over $J$ (a 100 K perturbation on
a 5000 K background), the 6\% error in the damping rate (in $J-S$)
results from a 0.5\% imprecision in the calculation of $J$ by
quadrature integration of Eq.\ \ref{eqint}. This is more than
sufficient for the SL9 problem, because $S\gg J$ in the SL9 shocks.

\subsection{Test 4: Radiative Equilibrium}
\label{ssect4re}

\citet{mihalas78} analytically derives the temperature {\em vs.}\
optical depth structure, $T(\tau)$, of a gray radiative equlibrium
atmosphere as a function of $T_\mathrm{eff}$.  Our modified code can
reproduce this structure.  Since the time scale to achieve radiative
equilibrium increases greatly with decreasing effective temperature,
we conducted this test using $T_\mathrm{eff}=5800$ K, a solar
temperature.  We started with a 2D volume of fluid having a constant
temperature (arbitrarily set to $6000$ K), and we added a large
arbitrary velocity perturbation in both dimensions.  After ${4.6
\times 10^5}$ sec of simultation time the fluid had converged to a
near-equilibrium state.  All state variables were nearly uniform in
the $r$ dimension, and $\rho$ and $p$ satisfied hydrostatic balance to
within a few parts in $10^7$.  The $T(\tau)$ relation was closely
equal to the analytic formula for a gray radiative equilibrium
atmosphere.  Figure \ref{figtest} shows $T(\tau)$ from the code
(points) versus the analytic formula (line).  The largest error is
$20$ K, and most layers are within 10 K (about 0.2\% error).  

\subsection{Test 5: Advection}
\label{ssect5adv}

ZEUS-3D uses \citet{vanLeer77} advection.  As noted earlier, we have
added an additional van Leer advection calculation specifically to
keep track of the plume portion of the total fluid.  We tested this
calculation by tailoring our algorithm to track the {\em total} mass
column density at each $r$, with a large oscillation in the $r$
velocity. The oscillation amplitude was 10 km sec$^{-1}$ at the top of
the model, decreasing as $(1+\tau^{2})$.  The amplitude was sinusoidal
in $r$, with a wavelength of 10 grid zones. This oscillation provides
a greater degree of fluid advection than occurs during the splashback,
so it is a stringent test of our van Leer algorithm.  Since our
algorithm is tracking the same fluid as the original ZEUS algorithm in
this case, the two calculations should track closely.  After 5000 sec
of simulation time, the largest difference between the column
densities from the two independent calculations was 0.5\%.  This
accuracy is more than sufficient to track the plume column densities
for the time scales of interest here.
 
\section{Model Results}
\label{secmr}

\subsection{Input Parameters and Nominal Plume}
\label{ssecmripnp}

Table \ref{tabpar} gives the parameters adopted for our nominal model.
Constraints on the ratio of plume to impactor mass come from
conservation of energy and the adopted velocity distribution used in
the ballistic Monte-Carlo model.  Both plume and impactor mass were
varied by the same scale factor in the ZEUS modeling (e.g., compare R
impact to L impact).

\begin{deluxetable}{lr}
\tablecaption{Model Parameters\label{tabpar}}
\tablewidth{0pt}
\tablehead{
\colhead{Parameter} &
\colhead{Value}}
\startdata
R-fragment mass & $1.7 \times 10^{14}$ g \\
L-fragment mass & $1.7 \times 10^{15}$ g \\
R-plume mass & $1.3 \times 10^{15}$ g \\
L-plume mass & $1.3 \times 10^{16}$ g \\
Impactor molecular weight & 18.0 g mole$^{-1}$\\
Jovian molecular weight & 2.28 g mole$^{-1}$\\
Impactor opacity & 10 cm$^{2}$ g$^{-1}$ \\
Jovian opacity & $8 \times 10^{-3}$ cm$^{2}$ g$^{-1}$\\
Ratio of specific heats, $\gamma$ & 1.2
\enddata
\end{deluxetable}

\citet{alexander94} calculate Rosseland mean opacities for solar system
matter.  At typical splashback temperatures ($600-2000$ K) the
Rosseland mean varies from 2.5 - 0.02 cm$^2$ g$^{-1}$. From the
composition tabulated by \citet{jessberger91}, the comet's metallicity
is enhanced by a factor of 40 over the sun.  Hence the opacity for the
impactor material will be in the range 100 - 1 cm$^2$ g$^{-1}$.  We
performed exploratory ZEUS calculations with opacity calculated as a
function of temperature at each time step, by interpolating in the
tabulated values given by \citet{alexander94}.  We find that this does
not dramatically improve agreement with observations.  We have
therefore used a constant intermediate value, 10 cm$^2$\ g$^{-1}$,
independent of temperature and density.

Because the plume is optically thin, varying the mass of the plume has
the same (nearly linear) effect on the modeled light curves as varying
the opacity.  Hence, the impactor mass and opacity do not change (to
first order) the {\em shapes} of our modeled light curves, they merely
scale the fluxes.  To specify a nominal plume mass (Table
\ref{tabpar}), we matched the peak of the 10-{\microns} modeled curve
to the flux of the well-observed R impact \citep{friedson95}.
Although we have relied on the R-impact to calibrate the mass scaling
factor for our models, most of our calculations used an L-fragment mass.

Our nominal plume contains a shell of mass at the highest velocity,
and the necessity for this `vanguard' of mass is discussed in paper I.
However, we have also calculated synthetic light curves for a plume
wherein the vanguard is eliminated, and a `non-vanguard' light curve
is shown in Sec.\ \ref{ssecmrnvlc}.  We adopted an observed
temperature profile for the jovian atmosphere derived from Voyager
IRIS measurements \citep{hanel79}, supplemented by Galileo probe
results \citep{seiff96} at the greatest heights.  Since this empirical
atmosphere is not in gray radiative equilibrium, coupling it to ZEUS
introduces unaccounted sources and sinks of energy, but these small
imbalances have no significant effect on our results.

How should we couple the plume from the expanding fireball into our
splashback model?  As the fireball expands to greater heights, it
encounters decreased atmospheric pressure, and its internal pressure
also decreases greatly.  Above some transition height, its motion
becomes controlled by ballistics rather than hydrodynamics.  Based on
the results of \citet{carlson97}, we infer this to occur at $z\sim400$ km
above the $1-$bar level.  Re-entry of the plume will shock the jovian
atmosphere even for $z>>400$ km, but these shocks at very great height
will produce negligible IR emission, and will not deaccelerate the
infalling plume significantly.  Accordingly, we add the infalling
plume mass and momentum fluxes to our model in the layer $400$ km
above the $1-$bar level, interpolating these fluxes from the
plume model at each ZEUS $r$ value and time step.

The infalling plume is believed to be cold (ZM95), so we have adopted
an initial plume temperature of 100 K, this being equal to the upper
atmospheric boundary temperature in gray radiative equilibrium. (Our
results are not sensitive to the initial plume temperature.)  The ZEUS
computational boundaries at maximum $z$ and $r$ were configured to be
transmitting.  The lowest-$z$ surface at the greatest atmospheric pressure
(5 bars) was specified as reflecting, although splashback effects do
not penetrate even close to this depth.  We experimented with a
variety of grid spacings in the model, in order to determine a grid
configuration which resolves the splashback shocks, while remaining
computationally tractable.  Our adopted grid uses 20 layers of 10 km
thickness in the region below 200 km, 100 layers of 4 km thickness
from 200 to 600 km (where the splashback effects are most prominent),
and an additional 40 layers of 10 km thickness above 600 km.  Our
radial grid spacing was $75$ km, extending from $r=0$ to $r=12000$ km
in 160 zones.  We verified that finer grid resolution or greater
extent will not change our results significantly.  

Because the plume is injected into the code at $z=400$ km, the layers
overlying this height participate only minimally by, for example,
`catching' matter which rebounds upward.  Accordingly, we do not
illustrate these layers in our Figures.  Note also that the matter
which they contain is counted as plume matter by the advection scheme
described in Sec.\ \ref{seccp}.  They initially contain jovian
atmosphere, but this accounting error has a negligible effect, since
the column density above $400$ km is very small.

Since our computations attempt specifically to isolate phenomena
related to the splashback, we ignore phenomena related to the $r=0$
boundary, and we couple to ZEUS only plume material ejected on paths
at or above the horizontal.  We also ignore the channel created by the
entry of the fragment into the atmosphere.  Some possible effects of
this channel are discussed in Sec.\ \ref{ssecmrop}.

\subsection{Shock Structure and Evolution}
\label{ssecmrsse}

The structure and evolution of splashback shocks has been discussed
by \citet[hereafter Z96]{zahnle96}.  Our results are
generally consistent with Z96's conclusions, but provide additional
insights.  Figure \ref{figtempvpre} illustrates our calculations of the shock
development and evolution at times extending up to the peak of the
main event at 700 seconds post-impact.  Each panel is labeled by time
and gives a false-color representation of the log of temperature, and
also includes overlaid velocity vectors.

At 100 sec, most of the infalling mass has not yet hit the shock and
is still cold, producing the dark region filling the upper left
corner. But the border of this region is bright, denoting a shock.
The shock is hottest at the right edge.  At this early time, the
infalling plume mass cannot exhibit large $z$ velocities, because
large $z$-velocity material requires longer times for re-entry.  But
large $r$ velocities are present, especially in the vanguard.  Since
this shock comes from the highest velocity material, it is hot, and is
most easily seen in the ascending branch of light curves at the
shortest wavelengths \citep{mcgregor96,ortiz97}, or in strong spectral
bands such as 3.3-{\microns} methane \citep{dinelli97}, rather than
the long thermal wavelengths \citep{lagage95,livengood95}.  In our
synthetic light curves it produces the `third precursors' noted by
\citet{mcgregor96} and other observers.  HST imaging of plumes at the
limb \citep{jessup00} shows emission attributed to an
upward-propagating shock near this time.  Upward-propagating shocks
from the early plume expansion are deliberately omitted from our
model, but the emission seen by \citet{jessup00} and the
radially-propagating shock modeled in Figure \ref{figtempvpre} are
parts of the same continuous process of plume expansion and
splashback.

At 300 seconds enough plume material has fallen to drive a shock
down to 200 km above the $1-$bar level (near $r=1500$ km), and the
lateral shock has now expanded to $r=4000$ km.  Additional infalling
plume material now encounters previously-fallen plume, and a second
shock begins to propagate back into the infall, as predicted by Z96.
This is becoming evident as a warm region above the hottest shock in
the $r=1000-3000$ km interval. Note also that for $r<1000$ km the
jovian atmosphere has temporarily rebounded from the pressure of the
initial plume infall.  This rebound is the first sign of a
pressure-mode oscillation which develops in the atmosphere at the
acoustic cutoff frequency.

At 700 seconds, near the peak of the observed (and our modeled) light
curves, the shock structure is complex.  For $r \lesssim 4000$ km the
temperature in the shocked plume rises with height.  This is due to
the existence and evolution of two shocks.  The lowest shock occurs
near the plume/atmosphere boundary, while the upper shock, due to
plume-on-plume collision, has now propagated up to nearly $z=400$
km. Note also that velocities in the lower shocked region are almost
entirely radial, because the $z$-component velocity has been
dissipated in the shock.  The reversed temperature gradient occurs
because the lower shock cools before the upper shock.  At any given
time, the lower shock has existed longer, so its cooling has proceeded
further.  Both shocks are optically thin, and independently cool by
radial expansion, as well as by radiation.  For a given total
velocity, plume mass having the smallest $z-$component and largest
$r-$component velocity will fall back first.  So the radial expansion of
the lower shock is greater than the upper, which contributes greatly
to its more rapid cooling.

The lower shock makes the dominant contribution to the IR light curves
at this time.  Although it is cooler, it has a much greater mass
density than the upper shock.  As time proceeds, regions of reversed
temperature gradient occur at increasingly larger $r$, and occupy much
larger area (in the $zr\theta$ model geometry, surface area is
proportional to $r^2$).  

Also at 700 sec, the lateral shock has moved to near $r=7000$\ km. Up
to this time, the expansion of the lateral shock is driven in part by
continuing infall of plume mass.  A cold rarefaction remains behind it
at high $z$ (the dark region at the top of the panel near $r=6000$
km).  A complex and interesting mass flow can be discerned in the
shock and rarefaction region, as revealed by the velocity vectors.
These vectors show the velocity of the {\em matter}, which is not the
same as the velocity of the shock itself.  Matter behind the
rarefaction (i.e., at smaller $r$) moves downward (from infall) as it
expands radially, reaching a minimum height of $z \sim 320$ km at the
same $r$ position as the rarefaction.  Ahead of the rarefaction (i.e.,
at greater $r$) there is a slight {\em upward} component to the
continued radial expansion. Some of this matter may fall back again at
later times, and has been postulated to account for the secondary
maxima in the light curves \citep{maclow96}.  However, as we will see
in Sec.\ \ref{ssecmrbounce}, another process is principally
responsible for the secondary maxima.

Figure \ref{figtempvpost} shows the structure and evolution of the
shocks at times after the peak of the main event.  At 1000 sec, the
reversed temperature gradient is accentuated by the collapse of the
vanguard.  The point where the vanguard intersects a given height
first begins to move radially inward at 720 sec.  The inward motion of
the intersection corresponds to ejection zenith angles decreasing from
45 -- 0 degrees, and this mass falls onto previously-impacted plume
matter.  A high velocity `plume-on-plume' collision is consistent with
the onset of high-excitation CO emission at 720 seconds, in excellent
accord with observations \citep{meadcr95}. This collapsing vanguard
produces a hot shock, which is visible near $r=2000-3000$ km.
Although the velocity of the matter in the vanguard is directed
radially outward, the shock produced by its collapse moves radially
inward with time.  This derives from the ballistics, wherein the last
portion of the vanguard to fall back is the portion ejected straight
upward, which (having no $r$ velocity) falls back at $r=0$.  As in the
case of the lateral shock (which is produced by the
horizontally-expanding portion of the vanguard), the shock from the
collapsing vanguard is hot, and produces its most significant effects
at the shortest wavelengths.  It is particularly prominent in our
synthetic light curves at 0.9 {\microns} (Sec.\
\ref{ssec09lc}), where it accounts for the flare observed to peak near
1000 seconds by several groups \citep[summarized by][]{ortiz97}.
Model temperatures in the flare shock reach $3000$ K.  The occurrence
of the highest shock temperatures in the collapsing vanguard is
qualitatively consistent with CO observations, wherein shock
temperature increases monotonically with time until $t \sim 1000$ sec
\citep{kim99}.  The collapse of the vanguard not only provides a
temperature sufficient to excite CO overtone transitions
\citep{noll97}, but also sufficient mass to provide significant
optical depth in these lines.  Consequently, the 2.3-{\microns} CO
features should appear in emission predominately {\em after} the peak
of the continuum light curves, in agreement with observations
\citep{meadcr95}.  It is not necessary to postulate dust formation
preferentially at specific times
\citep{hasegawa96} in order to explain the main characteristics of the
splashback.

At 1400 sec, infall of {\em plume} material has ceased, but radial
expansion of the fallen plume continues.  Collapse of the vanguard
leaves a low-pressure wake, which is accentuated by the continued
radial expansion.  This wake is filled by matter overlying the plume.
This in-rushing matter is quite noticeable on the 1400 sec panel,
especially in the $r=2000-8000$ km region, where it is seen falling
down and outward, joining the radial expansion of the collapsed plume.

In the model, the radial expansion or `sliding' of the plume is
unimpeded by molecular or turbulent viscosity, since viscosity terms
are not included in our `dry water' fluid equations.  Under this
condition, the plume gas will continue to expand until the layer
becomes infinitesimally thin in the $z$ dimension, but particulate
grains in the plume will drop below the shock to form a quasi-static debris
pattern in the atmosphere.  At 1400 seconds, the edge of the
radially-expanding plume continues to be marked by the lateral shock,
now near $r \sim 12,000$ km.  The radial expansion of this shock in
the model continues for much greater times, as it becomes essentially
a single propagating pressure pulse.  Its continued expansion accounts
for the rings of large radius observed by \citet[also Sec.\
\ref{ssecmr34r}]{mcgregor96}.

The 2000 sec panel shows that much of the upper atmosphere has cooled
significantly, and the level of continuum emission is greatly reduced
(i.e., the `main event' is over). However, even at much later times,
residual heating of the upper atmosphere for $z\ge 250$ km remains $\sim
200$ K, so that many spectral lines will still appear in emission
\citep{bezard97,kostiuk96,lellouch95}.  The lateral shock (now off the
panel) continues to propagate to larger radii, as discussed below.

Specific comments should be made concerning the agreement of the
modeled shock temperatures with observations.  Near maximum light, the
cooler portions of the modeled lower shock (600 K at $r=3000 $ km) are
in good agreement with the color temperatures seen in the IR continuum
\citep{nicholson95b}.  Regions at larger radius remain warmer than
observed (e.g., 1700 K at $r=5000 $ km), but are cooling rapidly.
Temperatures in the upper shock produced by the highest $z$-velocity
material at 1000 seconds (2500 K at $r=2000 $ km) are in good
agreement with spectroscopic observations in strong lines
\citep{knacke97}, but we cannot account for the 5000 K values reported
by \citet{kim99}.

\subsection{Light Curves}
\label{ssecmrlc}

The lack of wavelength-dependent opacities in our modeled light curves
limits them.  Nevertheless, using the wavelength-dependent Planck
function maintains sensitivity of shorter wavelengths to hotter
regions, thus accounting for much of the wavelength variation of light
curve morphology.  Our peak fluxes scale nearly linearly with opacity
and with plume mass, because the plume is optically thin.  Two factors
must be more solidly established in order to derive the impactor
masses from the observed light curves.  First, the fraction of impact
energy ($\eta$) which goes into lofting the plume must be known.
Second, accurate wavelength-dependent opacities must be incorporated
into the model.  At present, our adopted values ($\eta=0.3$ [ZM95] and
$\kappa=10$ cm$^2$ gm$^{-1}$) require an impactor mass for the
well-observed R fragment of $1.7 \times 10^{14}$ grams in order to
match the 10-{\microns} flux of \citet{friedson95}.  Assuming a mass
density of 0.5 g cm$^{-3}$, the implied diameter of the R fragment is
about 400 meters.  The mass of the largest fragment (L) would be an
order of magnitude larger, based on the 10-{\microns} flux observed by
\citet{lagage95}.

We discuss the comparison between the modeled and observed light
curves at several wavelengths spanning the range from just longward of
the visible to the thermal IR.  We plot the observed and modeled light
curves using scales (linear, log, etc.)  which follow the style of the
observed light curves as originally published.  Also, we have also
applied an approximate correction for viewing geometry to the
theoretical curves.  (This correction, while not negligible, is not
large enough to be a significant issue in the comparisons.)

\subsubsection{0.9-{\microns} Light Curves}
\label{ssec09lc}

Several observational groups observed large impacts (L and H) at 0.9
{\micronp}, using CCD detector arrays.  \citet{ortiz97} summarized the
observations.  \citet{schleicher94} obtained the most extensive data,
which are re-plotted in Fig.\ \ref{fig09lc}.  The observations as originally
published show a baseline drift toward higher intensities at the
latest times.  They attribute this baseline increase to greater
reflectivity of regions surrounding the splashback site, which is not
relevant in this context.  We therefore removed the observed baseline
using a low-order spline.

The observed light curves at this wavelength show a double-maximum
structure. The first maximum is broader, and peaks near 450 seconds,
corresponding to the main event.  The second maximum is much
sharper, and peaks near 1000 seconds.  The modeled light curve
exhibits both 450 and 1000 second maxima, whose relative widths
and amplitudes agree well with the observations.  However, the modeled
curve also exhibits an additional maximum near 250 sec, which is not
seen in the observations.

Radiation at this wavelength is particularly sensitive to the hottest
material in the splashback, because of the exponential term in the
Planck function.  Hot material is especially prominent at two epochs.
The first hot epoch begins quite early, as the portions of the
vanguard having small $z-$component and large $r-$component velocities
produce the lateral shock.  Methane observations show high
temperatures in this epoch persisting for much longer than can be
plausibly ascribed to the fireball phase \citep[e.g., Fig.\ 6 of][]{dinelli97}.
The first hot epoch involves the shock heating
of the jovian atmosphere as the lateral portions of the vanguard
expand.  A second hot epoch begins at 720 sec, caused by the collapse
of vanguard mass ejected at zenith angles less than 45 degrees. This
hot epoch is more familiar to observers; it involves infall onto
regions of previously-fallen plume material, which likely accounts for
the fact that it is quite prominent in CO \citep{meadcr95,kim99, mead01}.

The area encompassed by the lateral shock increases as the shock
expands, but its temperature decreases.  Hence its maximum
contribution to short-wavelength light curves occurs at a time
($200-300$ sec) which is determined by the balance between these
competing factors.  The shoulder on the Fig.\ \ref{fig09lc} modeled light curve
(marked `PC3' for `third precursor', see Sec.\ \ref{ssec2lc}) indicates the
maximum contribution of the lateral shock.  The observed light curve
does not show the PC3 feature, but it is similar to the model in that
it exhibits a significant rise almost immediately after impact, which
is not seen in observations at longer wavelengths \citep{nicholson96}.

The broad peak in the modeled and observed curves near 450 seconds
(marked `ME' for `main event') corresponds to the main event emission
seen at longer wavelengths, which are sensitive to a larger range of
temperatures.  However, both the model and observations show that the
main event at short wavelength is shifted significantly earlier in
time, i.e., $\sim450$ seconds post-impact {\em vs.}\ $\sim700$ seconds
at longer wavelengths.  This derives from the rapid cooling of the
lower shock, as discussed in Sec.\
\ref{ssecmrsse}.

The second hot epoch corresponds to the sharpest peak in the light
curve.  This feature has been called the `flare' by \citet{fitz96} and
is marked by `F' in Fig.\ \ref{fig09lc}.  We adopt this terminology
(note that the same term has been used more generally by some authors
to denote the entire period of post-impact brightening).  The origin
of the flare has not been adequately explained, although
\citet{ortiz97} correctly identified it with high-velocity ejecta.
Our modeled curve accounts well for its amplitude relative to the main
event, albeit with a slight error in the timing of the peak.  The
agreement is too good to be accidental, and we conclude that the model
has captured the essential physics of this feature.  The flare comes
from vanguard material launched on zenith angles less than
45{\degrees} and begins about 720 sec post-impact.  The last portions
of the vanguard to re-enter are those launched vertically; for a
vanguard velocity of 11.8 km sec$^{-1}$ these fall back at 1015 sec.
In the well-known HST limb images \citep[replotted in Paper
I]{hammel95}, the flare corresponds to the `plume flat on limb' phase
of plume collapse.  The modeled light curve declines very steeply
after this time, reflecting our adopted sharp velocity cutoff. The
observed flare occurs somewhat earlier, implying a slightly smaller
cutoff velocity for this impact.  The more gradual decline in the
observed flare also suggests that the velocity cutoff in the real
plume was not quite so sharp as in the model, i.e., the outer edge of
our modeled vanguard is too sharp.  Models which eliminate the vanguard
entirely do not produce an acceptable flare phase in their
0.9-{\microns} light curves.

\subsubsection{2-{\microns} Light Curves}
\label{ssec2lc}

Figure \ref{fig23lc} shows the light curve observed for the G impact by
\citet{mcgregor96}, plotted on a log scale in comparison to our
modeled light curve.  The morphologies of these light curves are very
similar to each other, again indicating that the model is accounting
to a large degree for the splashback physics. Essentially all features
of the observed light curve find counterparts in the model.  These are
marked PC3 (third precursor), ME (main event), F (flare), B (bounce),
and 2B (second bounce).  The PC3 feature was identified in the
observations by \citet{mcgregor96}.  The presence of third precursors on
the ascending branches of short-wavelength light curves is a natural
consequence of the expansion of the lateral shock, as discussed in
Sec.\ \ref{ssec09lc}.  In Fig.\ \ref{fig23lc}, the observed PC3
immediately follows the second precursor (from the fireball), which we
do not model with ZEUS.

Both the modeled and observed light curves exhibit a broad main event
phase between 500 and 1000 sec.  Both ME features are similar in that
they decline more sharply than they ascend.  This is due to the sudden
collapse of the vanguard in the flare phase; in the modeled light
curve this creates a particularly sharp drop in intensity near 1000
sec.  This bears an especially strong resemblence to observations of
the K impact \citep{mcgregor96,watanabe95}, but is unlike Keck
observations of the smaller R impact \citep{graham95}, which show a
smoother and more symmetric main event.  A tendency toward smoother
light curves for smaller impacts was previously noted by ZM95, and may
indicate the lack of a vanguard for smaller plumes (see Sec.\
\ref{ssecmrnvlc}).

Both light curves exhibit a secondary peak (the bounce) about 500--600 sec
after the main event peak.  Understanding this feature was the
original motivation for our work.  It is discussed in detail in
Sec.\ \ref{ssecmrbounce}.

There is one significant difference between the modeled and observed
light curves of Fig.\ \ref{fig23lc}.  The modeled curve decays at a
significantly greater overall rate following the main event peak.  To
some extent this may be due to saturation in the observed fluxes near
peak \citep{mcgregor96}, but it is also likely that the model cooling rate,
which occurs via radial expansion and radiative damping, is too large.
Since the model does not include viscous friction, the plumes expand
too rapidly in the model as compared to the observations (see Sec.\
\ref{ssecmr34r}).  Because radiation at this wavelength is quite
temperature-sensitive, a too-rapid rate of cooling will be quite
noticeable.  Inclusion of viscosity and wavelength-dependent opacities
in a future version of the model should allow us to refine the agreement
with observations.

\subsubsection{12-{\microns} Light Curves}
\label{ssecmr12lc}

Figure \ref{fig12lc} shows the light curve observed for the H impact by
\citet{lagage95}, in comparison to our model.  We again see good
agreement.  The predominant feature of thermal-IR light curves is their
relatively slow rate of descent following the main event peak.  This
produces an asymmetry wherein the rise to the main event occurs more
rapidly than the decline.  The model reproduces this behavior; the
intensity near 2000 seconds, relative to the peak intensity, is
very similar in the two curves. 

Two differences between the modeled and observed light curves are
noticeable on Fig.\ \ref{fig12lc}.  First, the main event is too broad in the
model relative to the observations.  Second, a weak flare feature
remains evident in the modeled light curve.  While the flare is a
dominant feature in both model and observations at the shortest
wavelengths (Fig.\ \ref{fig09lc}), it does not appear in the observations at 12
{\micronp}.  The modeled flare exhibits the same qualitative behavior,
being more prominent at shorter wavelengths, but does not fade
sufficiently at this long thermal wavelength; this discrepancy could
well derive from our crude gray opacity.

\subsection{The `Bounce'}
\label{ssecmrbounce}

Both the modeled and observed light curves in Figs. \ref{fig23lc} and
\ref{fig12lc} exhibit a secondary maximum, or bounce, at $\sim 600$
seconds following the main event ($\sim 1200-1400$ seconds
post-impact).  The modeled bounces are similar in amplitude to the
observations, and at least the first bounce occurs at the correct time
in the model.  Given the strong similarity between the model and
observations, we will look to the model to shed light on the physical
cause of this phenomenon.

We have investigated the nature of the modeled bounces extensively.
We extended the height of the top boundary in the models (as high as
$2000$ km), in order to be certain that material which rebounds upward
is not lost via model boundary conditions.  We used a variety of
jovian model atmospheres in order to evaluate the sensitivity to
atmospheric parameters.  We discuss several potential causes of this
phenomenon.

A widely-accepted explanation for the bounce (e.g., \citet{maclow96},
Z96) is matter deflected upward from the splashback shocks, falling
back ballistically at later times.  However, this matter is deflected
upward at relatively shallow angles, reminiscent of a stone skipped
across the surface of a pond, and its re-entry is not the principal
contributor to the bounce.  Instead, the principal cause of the
modeled bounce is seen on Fig.\ \ref{figtempvpost}, in the 1400 second
panel, the time of the bounce peak.  As noted in Sec.\
\ref{ssecmrsse}, the collapse of the vanguard leaves a low-pressure
wake which is accentuated by the radial expansion of the plume.
Matter rushes into this wake from the zones having $z\ge 400$ km.  As
this matter hits regions of increased density near $z\sim 300$ km, it
is also shocked and moderately heated.  If a reservoir of jovian air
were available above the collapsing plume, this mechanism would
explain not only the light curve bounces, but also the relative
prominence of methane and molecular hydrogen quadrupole emission in
the bounce phase \citep{mead01}.  However, the initial expansion of
the fireball may sweep aside, or entrain, much of the overlying jovian
atmosphere. Our current model cannot determine whether sufficient
jovian air remains above the plume to promptly fill the wake of the
vanguard collapse, since the answer depends on the physics of the
fireball/plume expansion.

Yet another `bounce mechanism' derives from the reaction of the
underlying atmosphere to the plume infall.  As noted in discussion of
Fig.\ \ref{figtempvpre}, the varying pressure of the overlying plume
induces an oscillation in the underlying atmosphere at the acoustic
cutoff period. This is a natural resonance of a stratified atmosphere
in pressure equilibrium \citep{lamb45}.  For a constant scale height
($H$) the acoustic cutoff period is $4\pi H /c$, proportional to
$T^{1/2}$, where $c$ is the sound speed and $T$ is temperature.  We
ran a series of computations using several isothermal jovian
atmospheric models, with different temperatures.  We verified that the
oscillation period in the model varies in proportion to $T^{1/2}$.
For our empirical model the resonance occurs at $\sim 450$
seconds. The acoustic mechanism contributes to the modeled bounce, but
determines the bounce period only after the first or second maximum.
Since the initial portion of the bounce phenomenon has a somewhat
longer period (500--600 sec), we expect that the period of the bounce
in observed light curves should shorten slightly with time.
\citet{nicholson95a} observed multiple bounces, but their observations
contain gaps which make it difficult to discern variations in the
bounce period.

\subsection{Radiative Damping}
\label{ssecmrrd}

The fact that the splashbacks were bright in infrared radiation
suggests that radiative emission may be an important, even dominant,
cooling mechanism.  Is the overall morphology of the light curves
determined by radiative cooling of the splashback regions?

Figure \ref{figradamp} shows the effect on the 2-{\microns} light
curve of turning off the radiative damping term in the model.  In the
absence of radiative damping, the peak flux is larger by a factor of 6
(0.8 in log flux).  Ignoring the obvious contradiction (that the
radiated flux is greater when we turn off radiation), we use this
comparison to evaluate the effects of radiative damping on the
splashback energy budget.  At 2000 seconds post impact, the flux in
the nominal model has fallen by two orders of magnitude, whereas the
undamped model shows a decrease of 0.6 in log flux (this being due to
radial expansion).  In the case of the 10-{\microns} light curve (not
illustrated), the nominal model falls by only 0.6 in log flux at 2000
sec, versus 0.3 for the undamped model.  So the effect of radiative
damping, relative to cooling by expansion, is greatest at the shortest
wavelengths.  ZM95 invoked an instantaneous balance between the
kinetic energy of the infalling plume and radiative losses.  Our
calculations confirm that this is a reasonable approximation.
However, the opposite approximation, where radiation is entirely
neglected and cooling occurs only by radial expansion, also produces
light curves whose {\it shapes} match the observations fairly well
(e.g., Fig.\ \ref{figradamp}, dashed line), especially for the longer
thermal wavelengths.

Radiative damping affects the time of maximum in the light curves
by a small, yet significant, amount.  Maximum of the nominal light
curve shifts slightly earlier than in the undamped case.  This is due
to the greater radiative losses from the hottest shocks, which occur
after 720 seconds. 

We monitored the modeled temperature profile of the atmosphere below
the depth of shock heating.  For the largest impacts (L, G, K) this
depth corresponds to $p\sim 0.5$ mbar, while for the moderate R
impact it is $p\sim 0.1$ mbar.  Atmospheric temperatures at greater
pressures did not vary significantly ($< 1$ K), indicating that heating
of the lower atmosphere by radiation from the overlying shocks was
negligible.  The {\it optical thinness} of SL9 shocks contrasts
with the larger terrestrial K/T impact, where radiation from the
splashback is believed to have ignited global terrestrial fires
\citep{melosh90}.

\subsection{Non-vanguard Light Curve}
\label{ssecmrnvlc}

Figure \ref{figradamp} also shows the 2-{\microns} light curve which
results from a plume wherein the vanguard is eliminated.  This causes
the main event to be more `rounded' and symmetric, whereas
observations of the G and K impact show main events which decline more
steeply than they ascend \citep{mcgregor96,watanabe95}.  The bounce is
broadened and delayed without the vanguard, and as noted in Sec.\
\ref{ssec09lc}, the vanguard is needed to match the flare seen at 0.9
{\micronp}.  Quite a few of the smaller fragments show symmetric main
events \citep[e.g.,][]{graham95}, and were not observed at 0.9
{\micronp}, so non-vanguard models might be preferred in those
instances.

\subsection{The 3--4-{\microns} Ring}
\label{ssecmr34r}

\citet{mcgregor96} describe the expansion of a large ring seen at
3--4 {\microns} but not at shorter or longer wavelengths.
Observations of the ring $\sim 4700$ and 7400 sec post-impact show
radii of $\sim 14,000$ and 18,000 km, respectively. The center of the
ring is offset from the impact location in the approximate direction
of the incoming fragment by 3600 km.  Extrapolating the ring radius at
the two observed times back to $t=0$ suggests an origin near $r=8000$
km from the impact site. Our model explains these effects, as discussed below.

The lateral shock produced by the vanguard accounts qualitatively for
the observed characteristics of this ring.  The shock produces a
relatively narrow, high-contrast feature, and remains sufficiently hot
to the large ring radii.  The initial velocity of the vanguard is 11.8
km sec$^{-1}$, which is much faster than the observed expansion
velocity of the ring.  However, the shock moves more slowly than the
velocity of the matter which matter which impacts it, the plume
mass being diverted upwards as it transfers only a fraction of its
momentum to the shock.  Also, the shock slows as it propagates to
larger radius, because the expansion does work against the ambient
atmosphere.  Linearly extrapolating two observations of ring position
back to $t=0$ may suggest $r=8000$ km as a starting position, but the
velocity is not constant, and the ring likely originated at $r=0$.

The offset between the ring center and the impact site is also
illusory.  Because the plume cone is tilted relative to vertical, the
lateral shock in the fragment-entry direction is driven by a larger
$r-$component velocity, and will expand faster than in the
anti-fragment direction.  This results in the apparent center of the
ring being substantially shifted toward the fragment direction.  We
calculated the shock positions and propagation speeds at larger radii
by running the code in a grid with $r$ extending to
30,000\ km.  In principle this calculation should be done in full 3-D
geometry at this radius, but that is beyond our current computational
capabilities.  We can closely approximate the results of a 3-D
calculation by extracting the infalling plume momentum and energy in
2-D `slices' in opposite radial directions from the impact point.
Figure \ref{figring} shows the ring radius and offset from these calculations,
compared to the observations by \citet{mcgregor96}.  The model
produces a radius which is too large by about 35\% compared to the
observations.  The offset, conversely, is too small in the model by
50\%.  The modeled ring velocity between the two observed times is
about 2.2 km sec$^{-1}$, versus an observed velocity of 1.4 km
sec$^{-1}$.

The expansion of the modeled ring is determined solely by the
non-viscous fluid Eqs.\ \ref{eqcont}--\ref{eqdsdtau}, whereas friction
from molecular and turbulent viscosity should also be included in
future models.  The addition of friction from this source can only
slow the modeled ring, and will improve the agreement with
observations.

The temperature of the modeled lateral shock at the large radii
corresponding to the ring observations is $\sim 600$ K.  This is
sufficient to produce thermal emission at the 3--4 {\microns}
wavelength of the observations.  The steep decline in the Planck
function makes such emission less detectable at significantly shorter
wavelengths, and optical thinness makes the emission faint at any
wavelength outside of strong spectral features.  \citet{mcgregor96}
discussed the peak wavelength of the ring emission, which they
concluded was extended longward of the 3.3-{\microns} methane band.
Also, the ring was not visible at 2.3 {\micronp}, although a methane
band occurs at that wavelength.  This has prompted investigation of
organic compounds such as tholins \citep{wilson97}.  But many
observations show (and our models agree) that the splashback took
place at quite low pressure levels (less than tens of ${\mu}$bar).
This insures that thermal emission from the ring was optically thin,
and most easily observed in {\em strong} bands of {\em abundant}
compounds such as the 3.3-{\microns} fundamental band of methane. In
this respect, note that the 2.3-{\microns} methane band has a strength
which is nearly two orders of magnitude less than the fundamental
\citep{pugh76}, so a 600 K methane ring should not be visible at 2.3
{\micronp}.  But the 3.3-{\microns} band will exhibit emission in
high-J transitions at longer wavelength.  For example, the P(18) and
P(19) methane lines near 3.5 {\microns} were observed for the C impact
by \citet{dinelli97}.  We therefore strongly suspect that the emission
of the 3--4-{\microns} ring was indeed due to methane in the 3.3
{\microns} band.

\subsection{Other Phenomena}
\label{ssecmrop}

As demonstrated in previous sections, our splashback model explains
and ties together many of the observed SL9 features.  Consequently, we
regard as significant the fact that it does not produce the
slowly-moving ($\sim 400$\ m sec$^{-1}$) dark rings which 
\citet{ingersoll95} attribute to gravity waves trapped in a deep water
layer.  Since the large oxygen abundance predicted by the
\citet{ingersoll95} model was not confirmed by the Galileo probe
\citep{niemann98}, the origin of these rings remains an open
question.  Recently, \citet{walterscheid00} produced outward-moving
gravity waves of the correct speed, using a hydrodynamic model wherein
energy is deposited in a cylindrical region with a radius from 250 to
1000 km, extending downward to pressures of $p\sim 0.2$ bars.  They
cite the plume splashback as the source of this energy deposition.

Our model allows gravity wave generation and propagation, but we find
no significant gravity waves excited by the splashback when energy and
momentum are coupled ballistically to the top layer of the atmospheric
model, as is correct purely for the splashback phase.  Splashback
effects do not penetrate even close to 0.2 bars, and we suggest that
the energy deposition adopted by \citet{walterscheid00} would be more
realistically ascribed to the entry of the fragment and/or the initial
expansion of the fireball.  The hot fireball provided a large source
of buoyancy which could potentially generate gravity waves at the
stratospheric level, and we would not see these gravity waves in our
current results.

\section{Summary}
\label{secsum}

We have modeled the interaction of SL9 ejecta plumes with the jovian
atmosphere during the splashback phase.  We treat the plumes as cold
matter in ballistic flight, using a Monte-Carlo method to populate the
\citet{ZM94} power-law velocity distribution.  Included in our nominal plume 
is a shell of mass near the high-velocity cutoff, which we call the vanguard.
Paper I optimized the plume parameters by comparing synthetic model
images with HST images of the impact sites.

In this paper, we have coupled the ballistic Monte-Carlo plumes to the
ZEUS-3D hydrocode, which we have modified to include radiative damping
in the gray approximation.  We validated the modified code by
comparing its output to analytic solutions for several test problems.
Using the code to follow the plume-atmosphere interaction has provided
new insights into the physics of a typical splashback.  The models
explain most of the major observed features of the SL9 light curves
and spectroscopy (especially for the larger impacts), including
several previously unexplained phenomena.  Continuum light curve
morphology from the model agrees with observations from 0.9--12
{\micronp}, without invoking {\em ad hoc} hypotheses such as dust
formation at specific times.

The plumes include matter ejected nearly horizontally.  The
horizontally-driven portion of the vanguard produces a lateral shock,
which at 200--300 sec post-impact produces hot methane emission as it
expands through the atmosphere.  Thus commences an early hot epoch in
the light curves, most noticeable at the shortest wavelengths.  The
third precursor phenomenon noted by \citet{mcgregor96} is due to this
shock, as is the hot methane emission observed by \citet{dinelli97},
and the rapid rise in light curves at 0.9 {\microns} \citep{ortiz97}.
Continued expansion gradually cools this shock, but it maintains a
ring-like morphology in the model.  This accounts for the rings seen
in 3--4-{\microns} images by \citet{mcgregor96}.  However, the radius
of our modeled ring is 35\% greater than the observed ring, and its
expansion velocity about two hours post-impact is 2.2\ km\ sec$^{-1}$
versus the observed velocity of 1.4 km sec$^{-1}$.  The future
inclusion of molecular and turbulent viscosity in the hydrocode will
slow the modeled expansion, bringing the modeled and observed ring
into closer agreement.

Shock temperatures and structure produced by the model agree well with
inferences from spectroscopic observations.  The first plume matter to
fall back produces a shock at the greatest pressures (typically tens
of ${\mu}$bars), with shock temperatures briefly approaching 2000 K.
Continued infall extends this to a second shock at greater heights,
while the lower shock cools rapidly by radiation and by its greater
radial expansion.  This more-rapid cooling of the lower shock accounts
for the bivariate nature of the shock temperatures derived from
observations \citep{nicholson96}.  Just after the peak of the light
curve, the lower shock has cooled sufficiently so that a fully
reversed temperature gradient is present. A second hot epoch commences
at this time (720 sec), with the infall of vanguard matter ejected at
zenith angles less than 45{\degrees}. At these ejection angles, the
infalling vanguard collides with previously-fallen plume matter.  This
`plume on plume' collision produces shock temperatures as high as 3000
K, and is responsible for the high-excitation CO emission seen at
these times.  The second hot epoch climaxes near 1000 sec, as portions
of the vanguard ejected at small zenith angles are the last to fall
back, and the HST limb images exhibit the `plume flat on limb' phase
of plume collapse. This produces the previously-unexplained sharp
spike in 0.9-{\microns} light curves, which \citet{fitz96} have termed
the flare.

Collapse of the vanguard creates a low-pressure wake, which is further
accentuated by the radial expansion of the plume.  If sufficient
jovian air is left in the immediate wake of the plume collapse, it
will rush into this wake, and be moderately shock heated,
producing hydrogen quadrupole emission, methane emission, and the
secondary light curve maximum called the bounce.  Although the timing
of the first bounce (500--600 seconds after the main event) is
initially determined by this mechanism, subsequent bounces in the
model occur at shorter periods (450 sec). This shortening of the
period occurs because the underlying atmosphere begins to oscillate at
the acoustic cutoff period.

Radiative damping contributes significantly to the cooling of the
heated plume, as does the radial expansion of the splashback regions.
Radiative damping is more rapid at the highest temperatures, so it
acts to limit temperatures of the hottest shocks, and is especially
important in shaping light curves at the shortest wavelengths.
Following the main event, residual heating in the upper atmosphere is
of order 200 K, in approximate agreement with observations
\citep{bezard97}.  Heating of the atmosphere below $\sim 1$ mbar, by
downward radiation from the overlying shocks, is negligible.

\acknowledgements

We thank K.\ Zahnle and M.-M. Mac\ Low\ for discussions relevant to
SL9 modeling, and G. Bjoraker for spectroscopic data.  P. Lagage,
T. Livengood, P. McGregor, P. Nicholson, J. Ortiz, and H. Schleicher
graciously sent us their data in digital form. We are grateful to an
anonymous referee, whose diligent review improved the presentation of
our results. A portion of this work was performed while J.\ H.\ held a
NAS/NRC Research Associateship at Goddard Space Flight Center.  This
research was supported by the NASA Planetary Atmospheres program.

\bibliography{slplume2-20-small}

\begin{figure}
\caption[]{\label{figgeom}Geometry used in the radiative-hydrodynamic model.}
\end{figure}

\begin{figure}
\ifthenelse{\lengthtest{\columnwidth = \textwidth}}{\epsscale{0.7}}{}
\ifthenelse{\lengthtest{\columnwidth = \textwidth}}{\epsscale{1}}{}
\caption[]{\label{figtest} a-c) Sod shock-tube test, using dimensionless
units. The initial values for $p$ and $\rho$ were both 1.0 on the left
of the computational boundary at $x=0.5$, and $10^{-4}$ and $10^{-1}$
respectively on the right of the boundary.  The boundary was removed
at $t=0$, and the figure illustrates the computational state at
$t=0.2$.  The points show the ZEUS-3D results, and the solid line is
the analytic solution calculated following \citet{hawley84}. d)
Radiative damping time scale test. The line represents the analytic
formula for the damping rate of temperature perturbations whose
spatial wavelength is $\tau_{\lambda}$ optical depths, and the points
are the results from the code. e) Radiative equilibrium test. The line
is the analytic temperature {\em vs.} optical depth relation for a
gray atmosphere having $T_\mathrm{eff}=5800$ K, and the points are the
results from the code.}
\end{figure}
 
\begin{figure}[p]
\ifthenelse{\lengthtest{\columnwidth = \textwidth}}{\epsscale{0.7}}{}
\ifthenelse{\lengthtest{\columnwidth = \textwidth}}{\epsscale{1}}{}
\caption[]{\label{figtempvpre}Temperature and velocity in the
splashback of a large impact (e.g., L) at three times leading to the
peak of the main event.}
\end{figure}

\begin{figure}[p]
\ifthenelse{\lengthtest{\columnwidth = \textwidth}}{\epsscale{0.7}}{}
\ifthenelse{\lengthtest{\columnwidth = \textwidth}}{\epsscale{1}}{}
\caption[]{Temperature and velocity in the splashback of a large
impact (e.g., L) at three times after the peak of the main
event.\label{figtempvpost}}
\end{figure}

\begin{figure}[tbp] 
\ifthenelse{\lengthtest{\columnwidth = \textwidth}}{\epsscale{0.95}}{}
\ifthenelse{\lengthtest{\columnwidth = \textwidth}}{\epsscale{1}}{}
\caption[]{\label{fig09lc}Observed light curve (dotted line) of the L
impact at 0.9 {\micronp}, from \citet{schleicher94}, in comparison to
our modeled light curve at this wavelength (solid line). The modeled
curve has been normalized in intensity to equal the observed curve at
450 sec. The notations PC3, ME and F stand for third precursor, main
event, and flare, respectively.}  
\end{figure}

\begin{figure}[tbp] 
\ifthenelse{\lengthtest{\columnwidth = \textwidth}}{\epsscale{0.94}}{}
\ifthenelse{\lengthtest{\columnwidth = \textwidth}}{\epsscale{1}}{}
\caption[]{\label{fig23lc}Observed light curve (lower plot) of the G
impact at 2.3 {\micronp}, from \citet{mcgregor96}, in comparison to
our modeled light curve at this wavelength.  The modeled curve has
been offset upward for clarity.  The notations PC3, ME, F,
and B, and 2B stand for third precursor, main event, flare, bounce,
and second bounce, respectively.}  
\end{figure}

\begin{figure}[tbp] 
\ifthenelse{\lengthtest{\columnwidth = \textwidth}}{\epsscale{0.97}}{}
\ifthenelse{\lengthtest{\columnwidth = \textwidth}}{\epsscale{1}}{}
\caption[]{\label{fig12lc}Observed light curve (lower plot, with
points) of the H impact at 12 {\micronp}, from \citet{lagage95}, in
comparison to our modeled light curve, offset in the ordinate for
clarity.  The `F' marks a residual flare feature in the modeled curve,
and `B' marks the first bounce.}
\end{figure}

\begin{figure}[tbp]
\ifthenelse{\lengthtest{\columnwidth = \textwidth}}{\epsscale{0.94}}{}
\ifthenelse{\lengthtest{\columnwidth = \textwidth}}{\epsscale{1}}{}
\caption[]{\label{figradamp}Effect of radiative damping, and omission of the
vanguard, on light curves for a large impact (e.g., L) at
2 {\micronp}.  The solid curve is the nominal plume model, including
radiative damping and a vanguard.  The dashed line shows the effect of
turning off radiative damping, and the dotted line shows the effect of
eliminating the vanguard.}
\end{figure}

\begin{figure}[tbp]
\caption[]{\label{figring}Observed radii (upper points) and offset (lower points) of the 3--4-{\microns} ring from G and K impact data by \citet{mcgregor96},
in comparison to the modeled radii (upper curve) and offset (lower
curve) versus time.}
\end{figure}

\end{document}